\documentstyle[preprint,aps]{revtex}
\tighten
\draft
\begin{document}
\widetext

\preprint{HUTP--97/A043}
\bigskip
\bigskip
\title{Global Symmetries in Duals of Supersymmetric SU(N)$\times$SU(M)\\
 and Application to Composite Axion}
\medskip
\author{Jihn E. Kim$^{1,2,}$\footnote{E-mail: 
jekim@huhepl.harvard.edu} and Bookyung Sul$^2$}

\vskip 1cm

\address{$^1$Lyman Laboratory of Physics, Harvard University, Cambridge, 
MA 02138\\
$^2$Department of Physics, Seoul National University, Seoul 151-742,
Korea\\
}
\date{\today}
\bigskip
\medskip
\maketitle

\begin{abstract}
{}In this paper, we construct a supersymmetric composite axion
({\it c-axion}) model based on the gauge group $SU(N)\times SU(M)$,
which is one possible physical application of the N=1 duality.
The dual of SU(M) is interpreted as the color gauge group.
We illustrate the existence of {\it c-axion} for the case of 
one dual quark in the dual gauge group $SU(\tilde M)$.  
\end{abstract}

\noindent {\bf 1. Introduction}
\vskip 0.4cm

Sometime ago the composite axion idea was suggested \cite{kim} so that
it is  the minimal extension of the standard model at low energy, i.e. 
the addition of just one pseudoscalar field $a$. In this composite
axion ({\it $\lq\lq$c-axion"}) model, one introduces an additional 
confining gauge group at the scale $\Lambda_h\sim 10^{12}$ GeV. In
supersymmetric theories, this confining gauge group can be the hidden 
sector gauge group needed for supersymmetry breaking.

Generally, the supersymmetric extension of {\it c-axion} 
is very much constrained
because one needs a superfield, transforming nontrivially both in
SU(3)$_c$ and in the hidden sector gauge group SU(N), which will be called 
the $Q$ field. $Q$ contributes positively to the QCD $\beta$-function,
which can lead to phenomenological difficulties such as in $\sin^2\theta_w$
and $\alpha_c$. But the most urgent requirement
for $Q$ is that it does not destroy the asymptotic freedom of QCD
below the scale $\Lambda_h$. Here, we impose this asymptotic freedom
condition for QCD. Indeed, {\it c-axion } in a supersymmetric
preon model has been considered in the literature \cite{pati}.
Dynamical symmetry breaking with strong gauge groups is an old idea
under the name of $\lq\lq$moose" \cite{moose}, but it was not
possible to draw exact results without supersymmetry. Supersymmetric
gauge theories offer more accurate predictions.

\def\M{$\tilde {\rm M}$}
In this paper, we study a possibility that the supersymmetric 
QCD is a dual theory of a supersymmetric SU(M) gauge theory.
It shown that a {\it c-axion} is generated in this dual
description, but our {\it c-axion} is different from that 
considered before in Ref. \cite{pati}. 

For the number ($N_f$) of flavors in SU(M) between
$M+1$ and $3M$, there can be a dual theory \cite{seiberg,others}.  We 
consider the possibility that the dual description is more appropriate
below a scale $\Lambda$. For $N_f$ quarks $Q_{\dot\alpha I}$ and $N_f$   
anti-quarks $\bar Q^{\dot\alpha}_I$ with I = $1,2,\cdots,N_f$,
the standard global charge assignment is
\begin{eqnarray}
  &\ \ \  V\ \ \   A\ \ \  R\nonumber\\
Q_{\dot\alpha I}\ \  &\ \ \ 1\ \ \ \  1\ \ \ \  1\\
\bar Q^{\dot\alpha}_I\ \ & -1\ \ \ \  1\ \ \ \ 1\nonumber
\end{eqnarray}
from which SU($N_f$)$_L\times$SU($N_f$)$_R$ and
two anomaly free U(1)'s, U(1)$_V$ and U(1)$_{\tilde R}$,
survive below $\Lambda$, where 
\begin{equation}
\tilde R=R-{M\over N_f}A.
\end{equation}
We will use this observation throughout the paper.

We assume that QCD, the dual description of SU(M), has a
perturbative coupling constant below the SU(M) scale 
$\Lambda_2$ and also it is
asymptotically free below the SU(N) confinement scale $\bar\Lambda_1$.  
For {\it c-axion},
one must study the dual of SU(N)$\times$SU(M) where we interpret
SU(N) as the confining force at the hidden sector scale and SU(M) as
the source of QCD. The general properties of SU(N)$\times$SU(M)
has been studied by Poppitz, Shadmi and Trivedi (PST) \cite{pst}.
The PST study, however, does not include the quarks of the standard
model, and does not attempt to apply to particle interactions.

\vskip 0.4cm
\noindent {\bf 2. Fundamental fields in SU(N)$\times$SU(M)}
\vskip 0.4cm

There are several scales involved in an SU(N)$\times$SU(M) gauge
theory. Let us denote the SU(N) scale as $\Lambda_1$ and the SU(M)
scale as $\Lambda_2$. Following PST, we consider the dual SU(\M)
of SU(M) gauge group. We take the viewpoint that it is more appropriate 
to describe in the dual theory SU(\M) below the SU(M) 
scale $\Lambda_2$. Thus below $\Lambda_2$, the spectrum of the theory is 
SU(\M) quarks $q$ and $\bar q$, and SU(M) singlet mesons $M$ and 
leptons $\bar L$.\footnote{ The SU(M) singlet leptons can transform 
nontrivially under SU(N).} 
The scale defining SU(M) singlet mesons $M$ will be described as
$\mu$. In the dual description, the SU(\M) scale $\bar
\Lambda_2$ is related to $\mu$ and $\Lambda_2$ by \cite{int,pst}
\begin{equation}
\Lambda_2^{b_0^{M}}\bar\Lambda_2^{b^{\tilde M}_0}=\mu^{b_0^M
+b_0^{\tilde M}}
\end{equation}
where $b_0^M=3M-N_f$ and $b_0^{\tilde M}=3\tilde M-N_f$ 
are the coefficient of the $\beta$ functions of the SU(M) 
and SU(\M), respectively.  $N_f$ is the number of 
flavors in SU(M) and SU(\M). Below $\Lambda_2$, the
number of fields transforming nontrivially 
under SU(N) is changed from that of above $\Lambda_2$ and
hence the SU(N) scale must be redefined as $\bar \Lambda_1$.
The $\bar\Lambda_1$ must be calculable in terms of $\Lambda_1$,
$\Lambda_2$ and $\mu$. These scales will be given after
presenting the model.

If we consider SU(N)$\times$SU(3), only two extra flavors
in addition to the three families of quarks are 
admissible.\footnote{Since we are interested in 
{\it c-axion} only, we do not pay attention to leptons.}
This is very restrictive. Therefore, a better scheme is to have an
SU(M) gauge theory above $\Lambda_2$ and QCD is interpreted
as the dual of SU(M) below $\Lambda_2$.  
The minimal representation for {\it c-axion} needs a representation
$(N,M)$ of SU(N)$\times$SU(M) \cite{kim}.  
In addition to the PST fields, we
introduce two more fields, the pre-up quark fields $R_P$,
and the anti-pre-up quark fields $\bar R_R$.
Thus we introduced N+1 flavors in SU(M).
These fields are shown in Table 1.

\vskip 0.5cm
\centerline{Table 1. {\it Global charges of the fields, including the 
$R$ charges of fermions.}} 
\centerline{\it Dots in the group representation columns 
denote singlets.}
\begin{center}
\begin{tabular}{|c|c|c|c|c|c|c|c|c|c|c|}
\hline
 & & & & & & & & & & \\
 Fields & $SU(N)$ &$SU(M)$&$L$& $C$ &$V$ & 
$A$ & $R$ & $\tilde R$ & $\tilde R_{fermion}$ & $Q_\mu$\\ 
\hline
 & & & & & & & & & & \\
$Q_{\alpha\dot\alpha} $ &$N$&$M$ & $0$ & $-1$
 &1 & $1$ & $1$ & $1-{M\over N_1}$ & $-{M\over N_1}$ & $\cdot$\\
 & & & & & & & & & & \\
$R_{\dot\alpha P}$ & $\cdot$ & $M$ & 0 & $N$ & 1 & 1 & 1 & 
 $1-{M\over N_1}$ & $-{M\over N_1}$ & $\cdot$ \\
 & & & & & & & & & & \\
$\bar R^{\dot\alpha}_T$ & $\cdot$ &$M^*$& $0$ &$0$ &
 $-1$ & $1$ & $1$ &$1-{M\over N_1}$ & $-{M\over N_1}$ & $\cdot$ \\
 & & & & & & & & & & \\
$\bar R^{\dot\alpha}_R$ &$\cdot$&$M^*$&$0$ &$0$ &
 $-1$ &$1$ &$1$ &$1-{M\over N_1}$ & $-{M\over N_1}$ &$\cdot$ \\
 & & & & & & & & & & \\
$\bar L^\alpha_A$ &$N^*$ &$\cdot$ &$-1$ &$0$&
 0& $1$ &$1$ &$1-{M\over N_1}$ &$-{M\over N_1}$ &$\cdot$\\
& & & & & & & & & & \\
$\lambda_N$ &$N^2-1$ &$\cdot$ &$0$ &0 &$0$ &$0$ &$\cdot$ &
 $\cdot$ &$1$ &$\cdot$\\
 & & & & & & & & & & \\
$\lambda_M$ &$\cdot$ &$M^2-1$ &$0$ &0 &$0$ &$0$ &$\cdot$ &
 $\cdot$ &$1$ &$\cdot$\\ 
 & & & & & & & & & & \\
\hline
\end{tabular}
\end{center}

\noindent 
Here the number of flavors $N_f$ is
\begin{equation}
N_1=N+1.
\end{equation} 
In Table 1, we have written $Q_\mu$ which is not a
relevant U(1) charge above the scale $\Lambda_2$.
It will be introduced below $\Lambda_2$. 
Its entries are represented as irrelevant dots. 
There are four types of chiral fields plus gauginos. 
(Note that the SU(M) gauge interactions cannot
distinguish $\bar R^{\dot\alpha}_T$ and $\bar R^{\dot\alpha}_R$.) 
Therefore, we considered five U(1) symmetries: L, C, V, A, and R.
The V, A, and R charges are given as in Eq. (1). For SU(M) quarks,
L charges are vanishing. The U(1)$_C$ is a subgroup of SU($N_1$)$_L$.
Note that we have not introduced the weak SU(2).
In this sense, this model is not realistic; 
but our objective here is to show the existence of {\it c-axion}. 

\vskip 0.4cm
\noindent{\bf 3. Global symmetries in SU(\M)}
\vskip 0.4cm

Suppose that $\Lambda_2$ is larger than $\Lambda_1$. Below $\Lambda_2$,
we can consider SU(M) singlet mesons $M$, baryons $B$ and
anti-baryons $\bar B$. For $N_f=M+1$ flavors of SU(M) theory, it has
been shown that it confines, leading to $N_f$ baryons $B_i$ and 
anti-baryons $\bar B^i$ where $i=1,2,\cdots,N_f$. 
For $N_f>M+1$, the SU(M) singlet baryon number increases very
rapidly.  
Seiberg \cite{seiberg} has shown that in this case it can be 
described by a dual gauge theory with the mesons $M$, dual quarks 
$q$ and anti-dual-quarks $\bar q$ in the gauge group SU(N$_f$--M). 
The baryons and anti-baryons are SU(N$_f$--M) singlets formed with the 
dual-quarks and anti-dual-quarks. In our case the dual gauge group
is SU($\tilde {\rm M}$), where
\begin{equation}  
\tilde M=N+1-M.
\end{equation}
The fundamental representation of the dual gauge group is $\tilde M$.
The $\tilde M$ fields are indexed by $\nu$. So the indices introduced
so far run up to

\begin{center}
\begin{tabular}{|c|}
\hline
 \\
\ \ \ $\alpha=1,2,\cdots,N$\ \ \ \\
 \\
\ \ \ $\dot\alpha=1,2,\cdots,M$\ \ \ \\
 \\
\ \ \ $\nu=1,2,\cdots,\tilde M$\ \ \ \\
 \\
\ \ \ $A=1,2,\cdots,M$\ \ \ \\
 \\
\ \ \ $T=1,2,\cdots,N$\ \ \ \\
 \\
\ \ \ $P,R=1$\ \ \ \\ 
 \\
\hline
\end{tabular}
\end{center}

\vskip 0.5cm

Thus for $N_f=N_1=N+1$, greater than $M+1$ but less than $3M$, we have
a dual description in the gauge group SU(\M). 
In this dual theory, the quarks
and anti-quarks, $q$ and $\bar q$, and SU(\M) singlet mesons $M$ 
will appear. These mesons are defined as d=1 superfields,
introducing a scale $\mu$,

\newpage 
\centerline{Table 2. {\it Global charges of the composite and elementary 
fields below $\Lambda_2$.}}
\centerline{\it Here $N_1=N+1$ and $M_{N1}=M+N+1$.}
\begin{center}
\begin{tabular}{|c|c|c|c|c|c|c|c|c|c|c|c|}
\hline
 & & & & & & & & & & & \\
 Fields &$SU(N)$&$SU(\tilde M)$&$L$& $C$ & $V$ &
 $Q_\mu$ & $\tilde R$ & $C_N$ & $V_N$ & $A_N$ & $R_N$ \\ 
\hline
 & & & & & & & & & & & \\
$\bar L^\alpha_A$ & $N^*$ & $\cdot$ & $-1$ & 0 & 0 & 0 & $1-{M\over N_1}$
 &--1 &--1 &1 &1 \\
 & & & & & & & & & & & \\
$q^\alpha_\nu $ &$N^*$&$\tilde M$& $0$ & 
${-M\over \tilde M}$ & ${M\over\tilde M}$ & $1$ & ${M\over N_1}$
 &$M\over\tilde M$ &--1 &1 &1 \\
 & & & & & & & & & & & \\
$M_{\alpha T}$ & $N$ & $\cdot$ & 0 & $-1$ & 0 & $-2$ &
 $2{\tilde M\over N_1}$ &0 &1 &1 &1  \\
 & & & & & & & & & & & \\
$M_{\alpha R}$ & $N$ & $\cdot$ & 0 & $-1$ & 0 & $-2$ &
 $2{\tilde M\over N_1}$ &0 &1 &1 &1 \\
 & & & & & & & & & & & \\
$r_\nu^P$ & $\cdot$ & $\tilde M$ & 0 & ${NM\over\tilde M}$ &
 ${M\over\tilde M}$ & 1 & ${M\over N_1}$
 & $M\over\tilde M$ &--1 &1 &1  \\
 & & & & & & & & & & & \\
$\bar r^{\nu T}$ &$\cdot$&$\tilde M^*$& $0$ &
$0$ & $-{M\over\tilde M}$ &$1$ & ${M\over N_1}$
 &$-{M\over\tilde M}$ &0 &--2 &0 \\
 & & & & & & & & & & & \\
$\bar r^{\nu R}$ &$\cdot$&$\tilde M^*$&$0$ &$0$ &$-{M\over \tilde M}$ &
 $1$ &${M\over N_1}$ & $-{M\over\tilde M}$ &0 &--2 &0 \\
 & & & & & & & & & & & \\
$M_{\alpha TP}$ &$\cdot$&$\cdot$& $0$ &$N$ & $0$ 
 & $-2$ & $2{\tilde M\over N_1}$ &0 &1 &1 &1 \\
 & & & & & & & & & & & \\
$M_{\alpha RP}$ &$\cdot$&$\cdot$&$0$ &$N$ &
 $0$ &$-2$ & $2{\tilde M\over N_1}$ &0 &1 &1 &1 \\
 & & & & & & & & & & & \\
$\lambda_N$ &$N^2-1$&$\cdot$ &0 &0 &0 &$0$ &1 &0 &0 &0 &1 \\
 & & & & & & & & & & & \\
$\lambda_{\tilde M}$ &$\cdot$&$\tilde M^2-1$&0 &0 &0 &$0$ &1
 &0 &0 &0 &1 \\
 & & & & & & & & & & & \\
\hline
 & & & & & & & & & & & \\ 
$\mu$ &$\cdot$&$\cdot$&0 &0 &0 & 2 &0 &$\cdot$ &$\cdot$ &
 $\cdot$ &$\cdot$ \\
 & & & & & & & & & & & \\
$\bar\Lambda_1^{2N-1}$ &$\cdot$&$\cdot$& $-M$ &$-M_{N1}$ &$M$ 
&--$M_{N1}$ & 2($N$--${M^2\over N_1}$) &0 &0 &2$N_1$ &2$N$ \\
 & & & & & & & & & & & \\
$\Lambda_1^{3N-M}$ &$\cdot$&$\cdot$& --$M$ & --$M$ &$M$ 
 &$0$ & 2($N$--${M^2\over N_1}$) &$\cdot$ &$\cdot$ &$\cdot$ &
 $\cdot$ \\
 & & & & & & & & & & & \\
$\Lambda_2^{3M-N-1}$ &$\cdot$&$\cdot$&0 &$0$ &0 &0 &0 &$\cdot$ 
 &$\cdot$ &$\cdot$ & $\cdot$\\
 & & & & & & & & & & & \\
$\bar\Lambda_2^{2N-3M+2}$ &$\cdot$&$\cdot$&0 &0 &
 0 &2$N_1$ & 0 & 0 & --$N_1$ & --$N_1$ & $N_1$--2$M$ \\
 & & & & & & & & & & & \\
\hline
\end{tabular}
\end{center}

\newpage
\begin{equation}
M={1\over\mu}(Q_{\alpha\dot\alpha}\ {\rm or}\ R_{\dot\alpha P})
\bar R^{\dot\alpha}_{T,R}.
\end{equation}
These composite particles together with the lepton $\bar L$
are shown in Table 2. 

With the U(1) charges, which will be given later, one can calculate
the charge of $\Lambda$ according to \cite{peskin}
\begin{equation}
\Lambda^{b_0}\propto e^{2\pi i\tau}\ \ {\rm where}\ \ 
\tau={\theta\over 2\pi}+i{4\pi\over g^2}.
\end{equation}
These are shown in the lower part of Table 2. Except for gauginos
the $\tilde R$ charges are for the bosonic partners.
We have not included the $A$ and $R$
charges in Table 2, because they are broken. The conserved U(1)
combination is
\begin{equation}
\tilde R=R-{M\over N_1}A.
\end{equation}
The parameter $\mu$ appears in the dual theory as given in Eqs. (3)
and (5) \cite{pst}. It is treated as a holomorphic variable.
Thus the meson fields $M$ introduces {\it an additional
global symmetry} U(1)$_\mu$. However, the conserved global symmetry
above and below $\Lambda_2$ must be four. Thus, Seiberg introduced
a superpotential of the form (dual quark)$\cdot$(meson)$\cdot$(dual
quark) \cite{seiberg}.
We will construct U(1) charges such that this kind of supperpotential
is generated at the scale $\Lambda_2$. To guarantee the 't Hooft
anomaly matching conditions
\cite{thooft}, we will represent the U(1) charges
in the form given in Eq. (1). Since we will be discussing SU(N)
confinement at $\bar\Lambda_1$, 
the U(1)'s must be of the form, U(1) subgroup of
SU($N_1$), U(1)$_V$, U(1)$_A$, and U(1)$_R$. Because 
these charges are relevant for the
SU(N) confinement, a subscript $N$ is added. Then the four
U(1) charges are
\begin{eqnarray}
&C_N=L+V \\
&V_N=L-{\tilde M\over 2M}V-{1\over 2}Q_\mu \\
&A_N=-L+{3\tilde M\over 2M}V-{1\over 2}Q_\mu \\
&R_N=\tilde R-{M\over N_1}L+{\tilde M\over 2M}V+\left(-{1\over 2}
+{\tilde M\over N_1}\right)Q_\mu
\end{eqnarray}
which are also shown in Table 2. Note that the following superpotenial
respects the four global symmetries and hence
is generated at the scale $\Lambda_2$,
\begin{equation}
W\sim q^\alpha_\nu (M_{\alpha T}\bar r^{\nu T}+M_{\alpha R}\bar r^{\nu R}).
\end{equation}

The dual quarks $q$ will form SU(\M) singlet baryons
which is identified as the baryons constructed from SU(M)
quarks $Q$ \cite{seiberg},
\begin{equation}
\epsilon^{\nu_1\nu_2\cdots\nu_{\tilde M}}q_{\nu_1}^{\alpha_{M+1}}
q_{\nu_2}^{\alpha_{M+2}}\cdots q_{\nu_{\tilde M}}^{\alpha_N}\sim 
\epsilon^{\alpha_1\cdots\alpha_M
\alpha_{M+1}\cdots\alpha_N}\epsilon^{\dot\alpha_1\cdots\dot
\alpha_M}Q_{\alpha_1\dot\alpha_1}\cdots Q_{\alpha_M\dot\alpha_M}
\end{equation}
whence we can schematically write $q$, for the purpose of
obtaining the U(1) charges, as
\begin{equation} 
q\sim {1\over \mu^{n_1}\Lambda_2^{n_2}}(QQ\cdots Q)^{1/\tilde M}
\end{equation}
where the number of $Q$'s are M. We considered only $\Lambda_2$
and $\mu$ in the above equation since we consider strong 
SU(M). The powers $n_1$ and $n_2$ are
determined to give dimension 1 superfield $q$ and appropriate
U(1) charges. From one column of Table 2, we obtain
\begin{equation}
n_1=-{1\over 2}\ ,\ \ {\rm and}\ \  n_2=-{1\over 2}+{M\over\tilde M}
\end{equation}
With this definition of $q$ we obtained all entries of the upper
part of Table 2. $\bar\Lambda_2$ is calculable from the matching
equation (3). $\bar\Lambda_1$ must be calculable in terms of
$\mu,\Lambda_1$, and $\Lambda_2$,
\begin{equation}
\bar\Lambda_1^{2N-1}=\Lambda_1^{3N-M}\Lambda_2^{{1\over 2}(3M-N-1)}
\mu^{-{1\over 2}(M+N+1)}.
\end{equation} 
All columns of Table 2 without dots satisfy the identification of 
$\bar\Lambda_1$ given in Eq. (17).  

The global symmetries are physical at the long distance scale,
which is culminated by the so-called 't Hooft's anomaly
matching condition \cite{thooft} which states that whatever
degrees of freedom one uses the global anomalies should match
to render the same physics at the long distance scale. If the
global anomalies do not match, the confinement is not obtained
or the global symmetry in question is broken by the strong
nonperturbative dynamics. 
The global symmetry\footnote{We can treat SU(N) as a global symmetry
for a moment as far as SU(M) gauge force is concerned.} below 
the scale $\Lambda_2$ is $SU(N_1)_L\times SU(N_1)_R\times 
U(1)_V\times U(1)_{\tilde R}$. Since U(1)$_C$ is a subgroup of
SU($N_1$)$_L$, all the above U(1)'s satisfy 't Hooft's anomaly
matching conditions.

For the generation of the composite axion, one needs a scale at
$\Lambda_2$. Most plausible scale is the gaugino condensation 
scale of the SU(M) gauginos,
\begin{equation}
<\lambda_2\cdot\lambda_2>\ne 0.
\end{equation}
Here we simply assume the gaugino condensation \cite{gaugino}.

We note from Table 2 that $\Lambda_2$ does not carry any global
quantum number; thus a dynamical scale provided by the SU(M)
confinement is $\Lambda_2$. 

Since we want to interpret SU(\M)
as QCD, we may have the condition $N-M=2$. 
The dual description is possible for $M+1<N+1<3M$, which
is satisfied for any $M$ and $N$ satisfying $N-M=2$.  

The scale $\mu$ must be comparable to $\Lambda_2$, since the mesons
will form at around $\Lambda_2$. Even though we keep track of $\mu$
and $\Lambda_2$, the magnitude of these scales will be set equal
in the end. So $\bar\Lambda_2$ is also comparable to $\Lambda_2$.
Similarly, $\bar\Lambda_1$ is comparable to $\Lambda_1$.

\vskip 0.4cm
\noindent {\bf 4. QCD and composite axion}
\vskip 0.4cm

The largest scale below $\mu$ and $\Lambda_2$ is supposed to be
$\bar\Lambda_1$. At $\bar\Lambda_1$, the SU(N) gauge group is 
assumed to become strong. For the SU(N) to become strong at
$\bar\Lambda_1$, the SU(N) $\beta$-function must be 
negative, which is always satisfied.
The SU(N) gauge theory given in Table 2 has $N+1$ flavors.
Thus it confines, forming SU(N) singlet baryons $l^A, \bar q^\nu
(\equiv\bar u^\nu), m^T,
m^R$, and SU(N) singlet mesons $K_{AT}, K_{AR}, N_{\nu T}, N_{\nu R}$.   
These mesons are also defined to be d=1 superfields, introducing
a scale $\mu_N$, similarly as we defined $M$'s in Eq. (6).
Below the scale $\bar\Lambda_1$, there survive three conserved U(1) 
charges. One anomalous U(1) is broken by the SU(N) instantons. 
These conserved charges are $C_N, V_N$ and $\tilde R_N$
which are shown in Table 3.

The superpotential given in Eq. (13) gives mass terms for
$N_{\nu T}, N_{\nu R}, \bar r^{\nu T}$ and $\bar r^{\nu R}$
at order $\bar\Lambda_1$. There remains only one flavor of
SU(\M) quark, $\bar u^\nu (\equiv \bar q^\nu), 
r_\nu^P$. Note that the surviving
quark pair is not $\bar r^{\nu R}$ and $r_\nu^P$.

The QCD  $\beta$ function between $\bar\Lambda_2$
and $\bar\Lambda_1$ is positive for $N>8$. But it is
negative below $\bar\Lambda_1$ due to the removal of
$N_{\nu T}, N_{\nu R}$ and their partners. We assume $N>8$.
We can equate $\alpha_{QCD}$(1 GeV)
and $\alpha_{\tilde M}$(10$^{12}$ GeV), i.e.
\begin{equation}
{1\over\alpha_{QCD}(\Lambda_{\rm QCD})}={1\over\alpha_{\tilde M}(
\bar\Lambda_2)}+{N-8\over 2\pi}\log{\bar\Lambda_2\over\bar\Lambda_1}
-{8\over 2\pi}\log{\bar\Lambda_1\over\Lambda_{\rm QCD}},
\end{equation}
implying
\begin{equation}
\Lambda_{\rm QCD}=\bar\Lambda_2\left({\bar\Lambda_1\over\bar\Lambda_2}
\right)^{N/8}\ \ ,\ \ \ {\rm for}\ \ N>8.
\end{equation}
Because we introduced only one quark, we will not draw any
phenomenological consequence from this relation. But we note that
a similar behavior of generating a small $\Lambda_{\rm QCD}$ can be
realized in a more realistic {\it c-axion} model.
By construction, the global symmetries shown in Table 3 do not
have SU(N) anomalies. Below $\bar\Lambda_1$, the SU(N) confines and
only SU(\M) nonabelian groups survive. All the three global
symmetries of Table 3 have SU(\M) anomalies. For the fields to
return to the original value after $2\pi$ phase rotation, the
U(1) quantum numbers must be integers. So the properly normalized
$C_N$ charges are obtained from those given in Table 3 multiplied
by $\tilde M$ and divided by the G.C.D. ($\equiv n_C$) of $M$
and $\tilde M$.
Similarly, the properly normalized $\tilde R$ charges are obtained
from the entries of Table 3 times $N_1$. Let the corresponding
properly normalized currents are $J^\mu_{C_N}, J^\mu_{V_N}$,
and $J^\mu_{\tilde R_N}$, which satisfy
\begin{eqnarray}
&\partial_\mu J^\mu_{C_N}={MN_1\over n_C}\{F\tilde F\} \nonumber\\
&\partial_\mu J^\mu_{V_N}=-N_1\{F\tilde F\} \\
&\partial_\mu J^\mu_{\tilde R_N}=N_1(2\tilde M-1)\{F\tilde F\} \nonumber
\end{eqnarray}
where $\{F\tilde F\}$ is the SU(\M) anomaly, $(1/2)\epsilon^{\mu\nu
\rho\sigma}F^a_{\mu\nu}F^a_{\rho\sigma}$. If the coefficient of this 
anomaly is greater than 2, there can be a potential domain wall
problem \cite{sikivie}. However, the calculation of the domain wall
number depends on the model \cite{rev}. Actually there exists only
one anomalous U(1) symmetry which is a combination of the above
three. The other two nonanomalous symmetries identify the different
vacua. If there is no massless quark, 
the number of degenerate vacua is given by \cite{choi}
\begin{equation}
N_{DW}={\rm G.C.D.}\ {\rm of}\ \ 
{MN_1\over n_C},\ N_1,\ N_1(2\tilde M-1)
\end{equation}
which is equal to the G.C.D. of $N_1$ and $MN_1/n_C$. Probably,
the inflation idea is necessary to remove the cosmologically
dangerous domain walls if $N_{DW}>1$.
But the gluino condensation $<\lambda\lambda>$ carries $2N_1$
units of $N_1\tilde R_N$ charge, gives the real domain wall
number of $N_{DW}/n_G$ with $n_G\equiv$ G.C.D of $N_{DW}$ and
$2N_1$. Since $N_1$ has been already considered in Eq. (22),
the real domain wall number is $N_{DW}$ for an odd $N_{DW}$
and $N_{DW}/2$ for an even $N_{DW}$.

\newpage
\centerline{Table 3. {\it SU(N) singlets. In the definition column,
the SU(N)}}
\centerline{\it quarks forming the baryons are also shown.}
\begin{center}
\begin{tabular}{|c|c|c|c|c|c|c|}
\hline
 & & & & & & \\
Fields&Definition & SU(\M) & $C_N$ & $V_N$ & $\tilde R_N$ &
$\tilde R_N(fermion)$\\  
\hline
 & & & & & & \\
$\bar u^\nu (\equiv\bar q^\nu)$ & $q^\alpha_\nu,\bar 
 L^\alpha_A$ & ${\tilde M}^*$ &
 ${MN\over\tilde M}$ & $-N$ & $N\over N_1$ & $-{1\over N_1}$ \\
 & & & & & & \\
$\bar l^A$ & $q^\alpha_\nu,\bar L^\alpha_A$ & {\bf 1} &$-N$ &$-N$ &
 $N\over N_1$ &$-{1\over N_1}$ \\
 & & & & & & \\
$m^T$ & $M_{\alpha T},M_{\alpha R}$ & {\bf 1} &
 $0$ & $N$ & ${N\over N_1}$ & $-{1\over N_1}$ \\
 & & & & & & \\
$m^R$ & $M_{\alpha T},M_{\alpha R}$ & {\bf 1} &
 0 & $N$ & $N\over N_1$ & $-{1\over N_1}$\\
 & & & & & & \\
$N_{\nu T}$ & ${1\over\mu_N}M_{\alpha T}q^\alpha_\nu$ & ${\tilde M}$ &
 $M\over\tilde M$ & 0 & $2\over N_1$ & $-1+{2\over N_1}$ \\
 & & & & & & \\
$N_{\nu R}$ & ${1\over\mu_N}M_{\alpha R}q^\alpha_\nu$ & $\tilde M$ &
 $M\over\tilde M$ & 0 & $2\over N_1$ & $-1+{2\over N_1}$ \\
 & & & & & & \\
$K_{AT}$ & ${1\over\mu_N}M_{\alpha T}\bar L^\alpha_A$ & {\bf 1} &
 --1 & 0 & $2\over N_1$ & $-1+{2\over N_1}$ \\
 & & & & & & \\
$K_{AR}$ & ${1\over\mu_N}M_{\alpha R}\bar L^\alpha_A$ & {\bf 1} &
 --1 & 0 & $2\over N_1$ & $-1+{2\over N_1}$ \\
 & & & & & & \\
\hline
 & & & & & & \\
$\bar r^{\nu T}$ & $\cdot$ & ${\tilde M}^*$ & $-{M\over\tilde M}$ &0 &
 ${2N\over N_1}$ &$-1+{2N\over N_1}$ \\
 & & & & & & \\
$\bar r^{\nu R}$ & $\cdot$ & ${\tilde M}^*$ &$-{M\over\tilde M}$ &
 0 & ${2N\over N_1}$ & $-1+{2N\over N_1}$ \\
 & & & & & & \\
$r^P_\nu$ & $\cdot$ & $\tilde M$ & $M\over\tilde M$ &--1 &$1\over N_1$ &
 $-1+{1\over N_1}$ \\
 & & & & & & \\
$M_{TP}$ & $\cdot$ &{\bf 1} &0 &1 &$1\over N_1$ &
 $-1+{1\over N_1}$ \\
 & & & & & & \\
$M_{RP}$ & $\cdot$ &{\bf 1} &0 &1 &$1\over N_1$ & 
 $-1+{1\over N_1}$\\
 & & & & & & \\
\hline
 & & & & & & \\
$\Lambda_{\rm QCD}^{3\tilde M-1}$ &$\cdot$ &$\cdot$ &
 ${MN_1\over \tilde M}$ &--$N_1$ &$\cdot$ &2$\tilde M$--1 \\
 & & & & & & \\
\hline
\end{tabular}
\end{center}

\vskip 0.4cm
\noindent {\bf 5. Yukawa terms and quark masses}
\vskip 0.4cm

Let us discuss the possibility of generating the light quark
mass. The superpotential generated at the SU(N) confining
scale is \cite{seiberg}
\begin{eqnarray}
&W_N=-{\mu_N^{N+1}\over\bar\Lambda_1^{2N-1}}
\epsilon^{A_1\cdots A_MR}
\epsilon^{\nu_1\cdots\nu_{\tilde M}}\epsilon^{t_1\cdots t_Nt_{N+1}}
K_{A_1t_1}\cdots K_{Rt_{M+1}}
\cdot N_{\nu_1t_{M+2}}\nonumber\\
&\cdots 
N_{\nu_{\tilde M-1}t_N}N_{\nu_{\tilde M}t_{N+1}}
+l^AK_{AT_1}m^{T_1}+\bar u^\nu N_{\nu T_1}m^{T1}
\end{eqnarray}
where $\{T_1\}=\{T,R\}=\{1,2,\cdots,N;R\}$ and $\{t_i\}=\{1,2,
\cdots;N,R\}$. There remains a massless quark with the above
superpotential. 

The global symmetries we considered so far are respected by the
gauge interactions. Any additional terms in the superpotential
would violate some of the global symmetries. But these additional
terms should be sufficiently small so that the exact results we
derived are intact. One possible superpotential we can add is
$vR_{\dot\alpha P}\bar R_R^{\dot\alpha}$ where $v\ll \mu$.
Another possible superpotential is $\epsilon (\mu_N/\mu)
\bar L^\alpha_AQ_{\alpha\dot\alpha}(\bar Q_{T_1}^{\dot\alpha})$.
For this not to interfere with the SU(N) strong force, we require
$\epsilon\ll 1$. One can also introduce nonrenormalizable
interactions of the form,
\begin{equation}
\sum_{T_1}{g\over M_{P\ell}}R_{\dot\alpha P}\bar R^{\dot\alpha}_{T_1},
\ \ \sum_A\sum_{T_1}{g^\prime\over M^3_{P\ell}}
\bar L_A^\alpha Q_{\alpha\dot\alpha}\bar Q^{\dot\alpha}_{T_1}
\bar L^\beta_AQ_{\beta\dot\beta}\bar Q^{\dot\beta}_{T_1},\ \cdots
\end{equation}
where $M_{P\ell}$ is the reduced Planck mass. Without the
nonrenormalizable interactions, one cannot give a mass to
$\bar u$. Just for the mass generation the first term is
enough. Then the full superpotential below the scale
$\bar\Lambda_1$ can be expressed in terms of SU(N) singlet
fields as
\begin{eqnarray}
&W=\mu_N\bar r^{\nu T_1}N_{\nu T_1}+r^P_\nu M_{PT_1}\bar r^{\nu 
T_1}-({\rm Det. \ term})
+l^AK_{AT_1}m^{T_1}+\bar u^\nu N_{\nu T_1}m^{T_1}\nonumber\\
&+v\mu M_{RP}+\sum_A\sum_{T_1}\epsilon\mu_N^2K_{AT_1}
+{g\mu^2\over M_{P\ell}}M_{PT_1}^2+{g^\prime\mu^2\mu_N^2\over
M_{P\ell}^3}K_{AT_1}^2
\end{eqnarray}
where the repeated indices imply summations. For supersymmetry,
we require for example,
\begin{eqnarray}
&{\partial W\over\partial l^A}=-\sum_T K_{AT}m^T-K_{AR}m^R=0\\
&{\partial W\over\partial m^R}=-\sum_Al^AK_{AR}-\bar u^\nu N_{\nu R}=0\\
&{\partial W\over\partial K_{AR}}=-l^Am^R+\epsilon\mu_N^2+{2g^\prime
 \mu^2\mu^2_N\over M_{P\ell}^3}K_{AR}=0\\
&{\partial W\over\partial M_{RP}}=r^P_\nu\bar r^{\nu R}+v\mu
 +{2g\mu^2\over M_{P\ell}}M_{RP}=0\\
&{\partial W\over\partial\bar r^{\nu R}}=\mu_NN_{\nu R}
+r_\nu^PM_{RP}=0.
\end{eqnarray}
Eqs. (29) and (30) determine
\begin{eqnarray}
&(I)\ M_{RP}=-{vM_{P\ell}\over 2g\mu},\ {\rm others=0},\ {\rm or\ }
\nonumber\\ 
&(II)\ N_{\nu R}=0,\ M_{RP}=0,\ <r_\nu^P\bar r^{\nu R}>=-v\mu.\nonumber
\end{eqnarray}
Eqs. (26) -- (28) determine
\begin{eqnarray}
&(A)\ K=0,\ <\bar u^\nu N_{\nu R}>=0,\ 
<l^Am^R>=\epsilon\mu_N^2,\ {\rm or}\nonumber\\
&(B)\ m^R=0,\ l^A=0,\ K=-{\epsilon M_{P\ell}^3\over 2g^\prime
\mu^2} \nonumber
\end{eqnarray}
where $K\equiv <K_{AT}>=<K_{AR}>$.  Solution (II) is expected
to break the color symmetry. Solution (B) has a huge 
vacuum expectation value for $K_{AR}$,
which may not be preferred in the inflationary scenario starting 
from the origin. Thus, we take Solution (IA).

Then the mass matrix for $(\bar u^\nu,\bar r^{\nu R})
\times (r_\nu^P,N_{\nu R})^T$ is, by taking $<l^A>=<m^R>$,
\begin{equation}
\left(\matrix{\ \ \ \ \ \ 0\ ,\ \epsilon^{1/2}\mu_N\cr 
-{vM_{P\ell}\over 2g\mu},\ \ \ \ \mu_N}\ \  \right).
\end{equation}
In the limit of $m_u\ll \mu_N$, we obtain
\begin{equation}
m_u\sim {\epsilon^{1/2}vM_{P\ell}\over 2g\mu}.
\end{equation}
Depending on the parameters, $m_u$ can be sufficiently small.
Thus the $u$ quark can be made massive,\footnote{In a real application,
it may be the $t$ quark mass.}
which is accomplished by the introduction of a small parameter $v$
which may be generated by a Higgs potential. Thus, the anomaly
considered in Eq. (21) cannot be rotated away by the redefinition 
of a massless quark.

Introduction of the superpotential destroys some of the
global symmetries. If all the global symmetries are broken
by the superpotential, then the {\it c-axion} may be too massive.
In our case, the global charges carried by the new terms are
multiples of those carried by $R_{\dot\alpha P}\bar R^{\dot
\alpha}_{T_1}\ ({\rm or\ }M_{RT_1})$ and $\bar L^\alpha_AQ_{\alpha
\dot\alpha}\bar Q^{\dot\alpha}_{T_1}\ ({\rm or\ }K_{AT_1})$.
Therefore, only two global symmetries are broken, leaving one
anomalous global symmetry $Q_{PQ}=\tilde R_N+(2/N_1)C_N-(1/N_1)V_N$.
For $M\ge 8$, the PQ symmetry violating operator $\epsilon_{\dot
\alpha_1\cdots\dot\alpha_M}\bar R^{\dot\alpha_1}_{t_1}\cdots
\bar R^{\dot\alpha_M}_{t_M}$ is sufficiently small, and  
the {\it c-axion} solves the strong CP problem.

We assumed that the Goldstone boson ({\it c-axion}) is
created by breaking $\tilde R$ by $<\lambda_2\cdot \lambda_2>$.
Then the axion scale $F_a$ is the U(1)$_{\tilde R}$ breaking scale,
\begin{equation}
F_a\sim\sqrt{<\lambda_2\cdot\lambda_2>^{2/3}
+<\lambda_1\cdot\lambda_1>^{2/3}}.
\end{equation}
This gaugino condensation may be realized by introducing supergravity
interactions. However, it is difficult to study the full supergravity
Lagrangian. On the other hand, we pretend to reproduce the symmetry
properties and its spontaneous breaking via an effective superpotential
in terms of gauge singlet fields.

The relevant superpotential is taken as \cite{k84}
\begin{equation}
W\sim (\lambda_1\Phi\Upsilon+m^2)S+(\lambda_2\Phi\Upsilon+m^2)
S^\prime
\end{equation}
where $\Phi, S$ and $S^\prime$ are gauge singlet fields and 
$\Upsilon$ is the effective chiral field for gaugino condensation. If the
couplings $\lambda_{1,2}$ are tuned to small values (arising
only through supergravity interaction), the discussion we
presented so far is almost intact. This superpotential is chosen so
that it preserves the $R$ symmetry with $S,S^\prime$ and $\Phi$
carrying 2, 2, and --2 units of $R$ charges, respectively, and also
it breaks supersymmetry through nonzero value of $\Upsilon$
and $\Phi$ for $\lambda_1\ne\lambda_2$. So it has the phenomenologically
anticipated properties of this paper. The supergravity 
interaction contains a term \cite{cremmer}
\begin{equation}
{1\over 4}M_{Pl}e^{-G/2}G^l(G^{-1})^k_lf^*_{\alpha\beta k}
\lambda^\alpha\lambda^\beta.
\end{equation}
With a gauge kinetic function $f\sim 1+a\Phi/\Lambda_2$, we
obtain the coupling $\sim \Phi\lambda\lambda$. This term can
arise from the cross terms in $|\partial W/\partial S|^2
+|\partial W/\partial S^\prime|^2$. Here, we
have not shown that $W$ results from supergravity, but mimick
some aspects of supergravity in terms of W.  Of course, Eq. (35)
is proportional to the gravitino mass, whence the overall coupling
in $W$ is proportional to the gravitino mass.  If we introduced
$W$ independent from the gaugino condensation, i.e. $\Upsilon$
is not the gaugino condensation, then the $R$ symmetry
breaking leads to a fundamental invisible axion \cite{inv}.

\vskip 0.4cm
\noindent {\bf 6. Conclusion}
\vskip 0.4cm

In conclusion, we exploited an interesting class of SU(N)$\times$SU(M)
models, attempting to interpret the known quarks as the dual 
particles, and studied the global symmetries and their breaking
by possible terms in the superpotential. There results a {\it c-axion}. 
For the extra factor group SU(N), we motivated
its rationale through generation of a {\it c-axion} which
require a representation of the type (N,M), implying a
strong force at the intermediate scale.
This intermediate scale can be the world
where the duality idea is physically applicable. 

\acknowledgments

{}We thank E. Poppitz for helpful communications.
One of us (JEK) thanks Howard Georgi for many helpful
discussions.  One of us (JEK) is
supported in part by the Hoam Foundation, the Distinguished
Scholar Exchange Program of Korea Research Foundation, SNU-Brown
exchange program and NSF--PHY 92--18167.


\bigskip
\begin{references}

\bibitem{kim} J. E. Kim, Phys. Rev. {\bf D31} (1985) 1733;
K. Choi and J. E. Kim, {\it ibid} {\bf D32} (1985) 1828.

\bibitem{pati} K. S. Babu, K. Choi, J. C. Pati, and X. Zhang,
Phys. Lett. {\bf B333} (1994) 364.

\bibitem{moose} H. Georgi, Phys. Lett. {\bf B151} (1985) 57;
H. Georgi, Nucl. Phys. {\bf B266} (1986) 274;
A. E. Nelson, Phys. Lett. {\bf B167} (1986) 200.

\bibitem{seiberg} N. Seiberg, Phys. Rev. {\bf D49} (1994) 6857;
Nucl.Phys. {\bf B435} (1995) 129.

\bibitem{others} K. Intriligator and N. Seiberg, Nucl. Phys. Proc.
Suppl. {\bf 45BC} (1996) 1;
K. Intriligator, N. Seiberg and S. Shenker, Phys. Lett. {\bf B342}
(1995) 152; 
M. Dine, A. Nelson, Y. Nir, and Y. Shirman,
Phys. Rev. {\bf D53} (1996) 2658; 
E. Poppitz and S. P. Trivedi, Phys. Lett. {\bf B365} (1996) 125; 
P. Pouliot, Phys. Lett. {\bf B367} (1996) 151; 
P. Pouliot and M. Strassler, Phys. Lett. {\bf B375} (1996) 175;
K. Intriligator and S. Thomas, Nucl. Phys. {\bf B473} (1996) 121;
C. Csaki, L. Randall and W. Skiba, Nucl. Phys. {\bf B479} (1996) 65.

\bibitem{pst} E. Poppitz, Y. Shadmi and S. P. Trivedi, Phys.
Lett. {\bf B388} (1996) 561; Nucl. Phys. {\bf B480} (1996) 125.

\bibitem{int} K. Intriligator and P. Pouliot, 
Phys. Lett. {\bf B353} (1995) 471.

\bibitem{peskin} M. E. Peskin, SLAC--PUB--7393, hep-th/9702094.

\bibitem{thooft} G. 't Hooft, in Cargese Summer Institute on {\it
Recent Developments in Gauge Theories}, ed. G. 't Hooft
(Plenum, 1980). 

\bibitem{gaugino} D. Amati, K. Konishi, Y. Meurice, G. C. Rossi,
and G. Veneziano, Phys. Rep. {\bf 162} (1988) 169, and 
references therein.

\bibitem{sikivie} P. Sikivie, Phys. Rev. Lett. {\bf 48} (1982) 1156.

\bibitem{rev} See, for example, J. E. Kim, Phys. Rep. {\bf 150}
(1987) 1.

\bibitem{choi} K. Choi and J. E. Kim, Phys. Rev. Lett. {\bf 55} 
(1985) 2637.

\bibitem{k84} A similar form was given in J. E. Kim,
Phys. Lett. {\bf B136} (1984) 378. 

\bibitem{cremmer} E. Cremmer et al, Nucl. Phys. {\bf B212} (1983)
413; E. Witten and J. Bagger, Phys. Lett. {\bf B115} (1982) 202.

\bibitem{inv} J. E. Kim, Phys. Rev. Lett. {\bf 43} (1979) 103;
M. A. Shifman, A. I. Vainshtein and V. I. Zakharov, Nucl. Phys.
{\bf B166} (1980) 493; M. Dine, W. Fischler and M. Srednicki,
Phys. Lett. {\bf B104} (1981) 199; A. R. Zhitnistkii, Yad.
Fiz. {\bf 31} (1980) 497 [Sov. J. Nucl. Phys. {\bf 31} (1980)
260]. 

\end{references}
\end{document}